\documentclass[aps,prl, twocolumn, superscriptaddress]{revtex4-1}

\usepackage{graphicx}
\usepackage{amssymb}
\usepackage[usenames]{color}
\usepackage{amsmath}
\usepackage{xspace}
\usepackage{longtable}
\usepackage{soul}
\usepackage{epstopdf}
\usepackage{array}
\usepackage{tabularx}
\usepackage{braket}

\begin{document}

\title{Hyperfine and spin-orbit coupling effects on decay of spin-valley states in a carbon nanotube}

\newcommand{\bea}{\begin{eqnarray*}}
\newcommand{\eea}{\end{eqnarray*}}
\newcommand{\bne}{\begin{equation*}}
\newcommand{\ede}{\end{equation*}}

\newcommand{\bnen}{\begin{equation}}
\newcommand{\eden}{\end{equation}}
\newcommand{\bean}{\begin{eqnarray}}
\newcommand{\eean}{\end{eqnarray}}
\newcommand{\bsen}{\begin{subequations}}
\newcommand{\esen}{\end{subequations}}

\newcommand{\bna}{\begin{array}}
\newcommand{\eda}{\end{array}}
\newcommand{\bnm}{\begin{enumerate}}
\newcommand{\edm}{\end{enumerate}}
\newcommand{\bni}{\begin{itemize}}
\newcommand{\edi}{\end{itemize}}

\newcommand{\VSD}{V_\text{SD}}
\newcommand{\ISD}{I}
\newcommand{\VGo}{V_\text{G1}}
\newcommand{\VGf}{V_\text{G4}}
\newcommand{\Vb}{V_\text{b}}
\newcommand{\gorb}{g_\text{orb}}
\newcommand{\Bvec}{\mathbf{B}}
\newcommand{\DeltaSO}{\Delta_\text{SO}} 
\newcommand{\DeltaKK}{\Delta_{KK'}} 
\newcommand{\tspin}{t_\text{spin}} 
\newcommand{\EN}{E_\text{N}} 
\newcommand{\Tstar}{T_2^*} 
\newcommand{\Techo}{T_\text{echo}} 
\newcommand{\Deltarms}{\Delta_\mathrm{rms}}
\newcommand{\tauE}{\tau_\mathrm{E}}
\newcommand{\tauS}{\tau_\mathrm{S}}
\newcommand{\Gammain}{\Gamma_\mathrm{in}}
\newcommand{\fR}{f_\mathrm{R}}
\newcommand{\Sg}{S_\mathrm{g}}
\author{T. Pei}
\email{tian.pei@materials.ox.ac.uk}
\affiliation{Department of Materials, University of Oxford, Parks Road, Oxford OX1 3PH, United Kingdom}

\author{A. P{\'{a}}lyi}
\affiliation{Department of Physics and MTA-BME Condensed Matter Research Group, Budapest University of Technology and Economics, Budapest, Hungary}

\author{M. Mergenthaler}
\affiliation{Department of Materials, University of Oxford, Parks Road, Oxford OX1 3PH, United Kingdom}

\author{N. Ares}
\affiliation{Department of Materials, University of Oxford, Parks Road, Oxford OX1 3PH, United Kingdom}

\author{A. Mavalankar}
\affiliation{Department of Materials, University of Oxford, Parks Road, Oxford OX1 3PH, United Kingdom}

\author{J. H. Warner}
\affiliation{Department of Materials, University of Oxford, Parks Road, Oxford OX1 3PH, United Kingdom}

\author{G. A. D. Briggs}
\affiliation{Department of Materials, University of Oxford, Parks Road, Oxford OX1 3PH, United Kingdom}

\author{E. A. Laird}
\email{edward.laird@materials.ox.ac.uk}
\affiliation{Department of Materials, University of Oxford, Parks Road, Oxford OX1 3PH, United Kingdom}

\begin{abstract}
The decay of spin-valley states is studied in a suspended carbon nanotube double quantum dot via leakage current in Pauli blockade and via dephasing and decoherence of a qubit. From the magnetic field dependence of the leakage current, hyperfine and spin-orbit contributions to relaxation from blocked to unblocked states are identified and explained quantitatively by means of a simple model. The observed qubit dephasing rate is consistent with the hyperfine coupling strength extracted from this model and inconsistent with dephasing from charge noise. However, the qubit coherence time, although longer than previously achieved, is probably still limited by charge noise in the device.
\end{abstract}
\maketitle

The co-existence in carbon nanotubes of spin and valley angular momenta  opens a host of possibilities for quantum information~\cite{Churchill2009Nat,Flensberg2010,Pei2012,Viennot2015}, coherent coupling to mechanics~\cite{Ohm2012, Palyi2012}, and on-chip entanglement~\cite{Braunecker2013, Hels2016}. Spin-orbit coupling~\cite{Kuemmeth2008} provides  electrical control, but introduces a relaxation channel. However, measurements of dephasing and decoherence~\cite{Churchill2009PRL,Laird2013,Laird2015} show that spin and valley qubit states couple surprisingly strongly to lattice nuclear spins and to uncontrolled electric fields, e.g. from thermal switchers. Realising these possibilities requires such effects to be mitigated. Here we study leakage current in a Pauli blockaded double quantum dot to identify spin-orbit and hyperfine contributions to spin-valley relaxation~\cite{Koppens2005,Nadj-Perge2010,Pei2012}. By suspending the nanotube, we decouple it from the substrate~\cite{Laird2013}. Measuring a spin-valley qubit defined in the double dot, we find dephasing and decoherence rates nearly independent of temperature, and show that charge noise cannot explain the observed dephasing, supporting the conclusion that despite the low density of $^{13}$C spins, hyperfine interaction causes rapid dephasing in nanotubes~\cite{Churchill2009PRL, Laird2013}.

The measured device [Fig.~1(a-b)] is a carbon nanotube suspended by stamping between two contacts and over five gate electrodes G1-G5~\cite{Wu2010, Pei2012, Mavalankar2016, Supp}. Gate voltages $V_\mathrm{G1}-V_\mathrm{G5}$, together with Schottky barriers at the contacts, define a double quantum dot potential. The dot potentials are predominantly controlled by gates G1 (for the left dot) and G4-5 (for the right dot), while the interdot tunnel barrier is controlled by gates G2-3. For fast manipulation, gates G1 and G5 are connected via tees to waveform generator outputs and a vector microwave source. The device is measured in a magnetic field $\Bvec =(B_X,B_Y,B_Z)$, with $Z$ chosen along the nanotube and $X$ normal to the substrate. Experiments were in a dilution refrigerator at 15~mK unless stated.

\begin{figure}[b]
\centering
\includegraphics{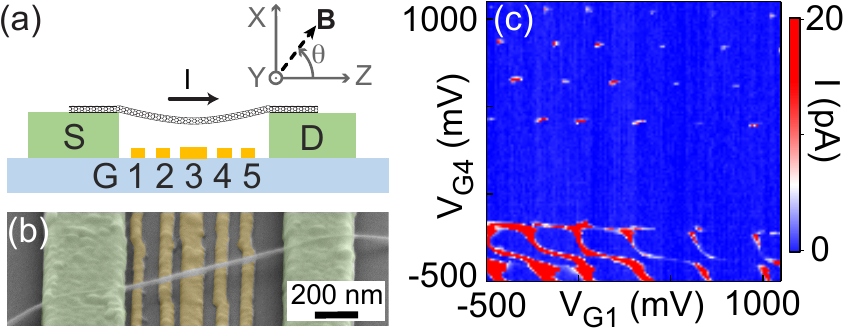}
\caption{(a) Schematic and (b) scanning electron microscopy image of a device lithographically identical to the one measured. The nanotube is suspended between contact electrodes (130~nm Cr/Au, marked S and D) and over gate electrodes (20~nm Cr/Au, marked G1-5) patterned on a Si/SiO$_2$ substrate. Field axes are indicated. For imaging, 2 nm of Pt was evaporated over this chip. (c) Current as a function of gate voltages $V_\text{G1}$ and $V_\text{G4}$, mapping out a double quantum dot stability diagram.}
\label{Fig1}
\end{figure}

To map charge configurations of the double quantum dot, we measure the current $\ISD$ through the nanotube with source-drain bias $\VSD=8$ mV applied between the contacts [Fig.~1(c)]. As a function of $V_\text{G1}$  and $V_\text{G4}$, the honeycomb Coulomb peak pattern is characteristic of a double quantum dot, with honeycomb vertices marking transitions between particular electron or hole occupations~\cite{VanderWiel2003}. A horizontal stripe of suppressed current around $V_\text{G4}=200$~mV indicates depletion of the right dot in this gate voltage range. The width of this stripe implies a band gap of 120~meV. No such suppression is observed as a function of $V_\text{G1}$, indicating that the left dot is doped across the entire range. Since at room temperature conductance decreases with increasing $V_\text{G1}$, we believe that the left dot is doped with holes, implying that p-p and p-n double-dot configurations are accessible~\cite{Steele2009}. Within each honeycomb region, we can therefore assign absolute electron or hole occupations to the right dot, but only relative hole occupations to the left dot.

Because tunnelling between quantum dots is governed by selection rules on spin and valley quantum numbers, transport through the device is subject to Pauli blockade~\cite{Ono2002}. This arises because the exclusion principle imposes an energy cost to populate spin-valley triplet states in a single quantum dot. Interdot tunnelling from a spin-valley triplet formed between the two dots is therefore blocked, suppressing $\ISD$ for certain gate and bias settings. In this blocked regime, a leakage current gives information about spin and valley relaxation.

We focus on Pauli-blockaded transport with the double dot tuned to a p-n configuration \cite{Churchill2009PRL,Pei2012}. Figure 2(a) shows $\ISD$  as a function of gate voltage near a $(n_h,1_e)\rightarrow((n+1)_h,2_e)$ transition. Here ($n_h$,$m_e$) denotes a configuration with $n_h$($m_e$) holes (electrons) in the left (right) dot. Two overlapping current triangles are seen, as expected for double-dot Coulomb blockade~\cite{VanderWiel2003}; in the lower triangle, transport occurs via the cycle of tunnelling events $((n+1)_h,1_e)\rightarrow(n_h,1_e)\rightarrow((n+1)_h,2_e)\rightarrow((n+1)_h,1_e)$, and in the upper triangle via $(n_h,2_e)\rightarrow(n_h,1_e)\rightarrow((n+1)_h,2_e)\rightarrow(n_h,2_e)$. The low current near the triangle baselines  is indicative of Pauli blockade suppressing the second step in each sequence, as expected for odd $n_h$.

To characterize the energy levels and spin-valley relaxation, we measure $\ISD$ as a function of magnetic field and double dot detuning $\varepsilon$, defined as the difference of electrochemical potential between left and right dots~\cite{Pei2012}. Detuning is swept by adjusting $\VGo$ and $\VGf$ along the diagonal axis marked in Fig.~2(a), with the triangle baselines marking $\varepsilon=0$. Figure~2 shows data as a function of magnetic field parallel [Fig.~2(b)] and perpendicular [Fig.~2(c)] to the nanotube, and as a function of field angle $\theta$ in the $XZ$ plane [Fig.~2(d)]. The triangle edge locations in gate voltage space give information about the double dot energy levels; the upper edges in Fig.~2(b-d)  correspond to ground state degeneracy ($\varepsilon=0$) between left and right dots, while the lower edge marks the degeneracy of the right dot ground state with the Fermi level in the right lead. From the evolution of the lower edge, which maps the energy of the two-electron state, we extract orbital $g$-factor $\gorb \approx 15$, spin-orbit coupling $\DeltaSO \approx 300~\mu$eV, and valley mixing parameter $|\DeltaKK| \lesssim 80~\mu$eV for the right dot, consistent with measurements on neighbouring transitions~\cite{Supp} and on other single-wall nanotube devices~\cite{Minot2004,Kuemmeth2008,Steele2013,Laird2015}. However, in similar measurements cutting through the left triangle edge and therefore tracking levels of the left dot, no clear field dependence was seen~\cite{Supp}. This is explained either by stronger valley mixing $\DeltaKK$ in the left dot (e.g. due to disorder) or by suppression of the valley magnetic moment by large hole occupation~\cite{Jespersen2011,Laird2015}. Thus the single-dot spectrum differs between left and right dots.

\begin{figure}[t]
\centering
\includegraphics[width=1\columnwidth]{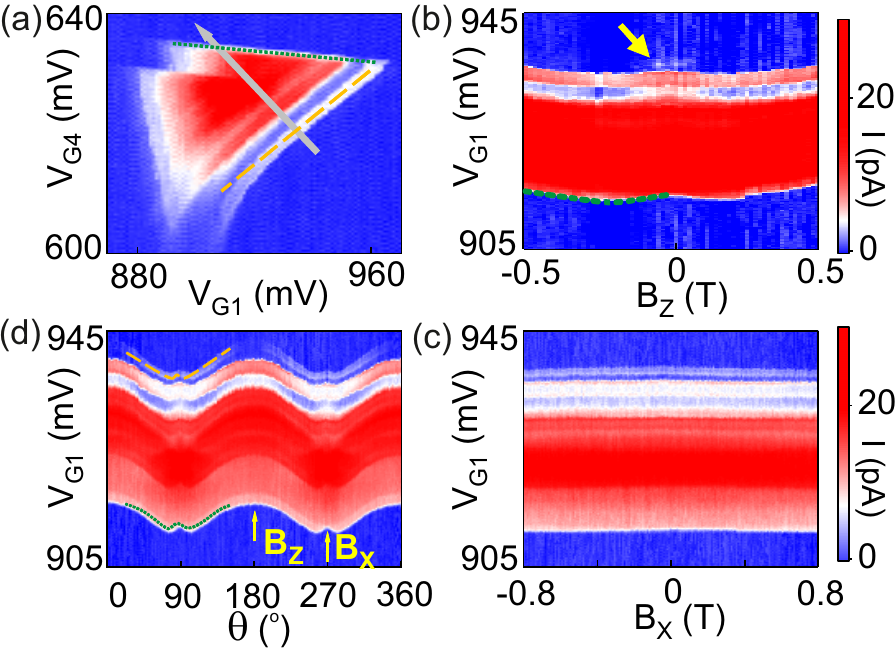}
\caption{(a) Current at a Pauli blocked transition, with $\VSD=8$~mV, $V_\text{G2}=V_\text{G3}=-210$~mV, $\Bvec=0$. Dashed (dotted) line marks ground-state degeneracy between left and right dots (between right dot and lead). Arrow marks detuning axis. (b) Current as a function of $\VGo$ along the detuning axis and of magnetic field parallel to the nanotube. Arrow marks a region of Pauli blockade leakage current near zero field. (c) As (b) for perpendicular field. (d) As a function of field angle for $|\Bvec| = 0.8$~T. Color scales in (a), (d) match (c). Dashed and dotted curves highlight the same transitions as in (a).}
\label{Fig2}
\end{figure}

We now study the field dependence of Pauli blockade leakage current to gain insight into spin-valley relaxation mechanisms~\cite{Koppens2005,Churchill2009PRL,Pei2012}. This leakage current is evident for small $\varepsilon$ (top of the current band) in Fig.~2(b-d), and shows a strong dependence on field direction.
As a function of magnetic field, the current is maximal around $B_Z=0$ [Fig.~2(b)], but varies only weakly with $B_X$~[Fig.~2(c)]. In fact, the leakage current can even show a dip at $B_X=0$~\cite{Supp}.
This different behavior is attributed to different complex phases of $\DeltaKK$ in the two dots, which in a perpendicular field lead to non-aligned effective Zeeman axes nearly independent of field strength~\cite{Szechenyi2013,Szechenyi2015} and therefore leads to leakage current nearly independent of $B_X$~\cite{Supp}. Similar behavior in some other systems \cite{schroer2011, Brauns2016, li2015pauli, Nadj-Perge2012} is due to anisotropy of the $g$-factor.

The low-$B_Z$ current peak is an indication of hyperfine-mediated relaxation. To study it in more detail, Fig.~3(a) shows measurements for different settings of the interdot tunnel barrier. Here the barrier is tuned by the voltage $\Vb \equiv V_\text{G2} = V_\text{G3}$. For a range of barrier settings, the central peak is accompanied by two side peaks. This contrasts with previous measurements in GaAs, InAs, and InSb, where a hyperfine-induced peak in Pauli blockade leakage current at zero field evolves to a double peak as tunnel coupling is increased \cite{ Koppens2005, Pfund2007, Nadj-Perge2010, Nadj-Perge2012}. Here, side peaks instead occur in conjunction with a zero-field peak.

This behaviour is explained by considering the effects of hyperfine interaction together with spin-orbit coupling. Consider the zero-field behavior first, and for concreteness focus on large detuning as shown in Fig.~3(b). The spin-valley degree of freedom associated with the unpaired particle in each dot forms an effective spin-$\frac{1}{2}$ Kramers doublet $\{\ket{\Uparrow},\ket{\Downarrow}\}$. Without hyperfine coupling and spin-dependent tunneling, the two-particle states are an effective $((n+1)_h,2_e)$ singlet ground state $\ket{\Sg}$, a singlet excited state
$\ket{S}\equiv \frac{1}{\sqrt{2}} \left(\ket{\Uparrow\Downarrow} - \ket{\Downarrow\Uparrow}\right)$, and $(n_h,1_e)$ triplet states $\ket{T_+}  \equiv \ket{\Uparrow\Uparrow}$,
$\ket{T_0} \equiv \frac{1}{\sqrt{2}} \left(\ket{\Uparrow\Downarrow} + \ket{\Downarrow\Uparrow}\right)$,
and $\ket{T_-} \equiv \ket{\Downarrow\Downarrow}$. In this effective spin basis, the main effect of introducing spin-orbit interaction is to cause spin-dependent tunnelling \cite{Danon2009, Klinovaja2011, Winkler2003}, described by Hamiltonian $H_\text{tun} = t \ket{S}\bra{S_g} + i t_\text{spin} \ket{T_u}\bra{S_g} + h.c.$, where $t$ and $t_\text{spin}$ are respectively the spin-conserving and spin-flip tunnel couplings. This Hamiltonian couples one superposition $\ket{T_u}$ of triplet states to the $((n+1)_h,2_e)$ singlet $\ket{S_g}$, while two orthogonal triplet superpositions $\ket{T_{b1}}$ and $\ket{T_{b2}}$ remain uncoupled. The energy eigenstates of the $(n_h,1_e)$ configuration are therefore $\ket{T_{b1}}$, $\ket{T_{b2}}$, $\ket{M_1}$ and $\ket{M_2}$, where $\ket{M_{1,2}}$ are mixtures of $\ket{S}$ and $\ket{T_u}$, and $\ket{M_2}$ remains degenerate with $\ket{T_{b1}}$ and $\ket{T_{b2}}$. Since $\ket{M_1}$ and $\ket{M_2}$ both have a finite $\ket{S}$ component, spin-independent inelastic interdot tunnelling processes (e.g. phonon-assisted tunnelling) allow charge relaxation into $((n+1)_h,2_e)$; however, the two uncoupled triplets $\ket{T_{b1}}$ and $\ket{T_{b2}}$ cannot relax in this way and therefore block the current. The spectrum, including magnetic field dependence, is shown in Fig.~3(b).
\begin{figure}[t]
	\centering
	\includegraphics[width=1\columnwidth]{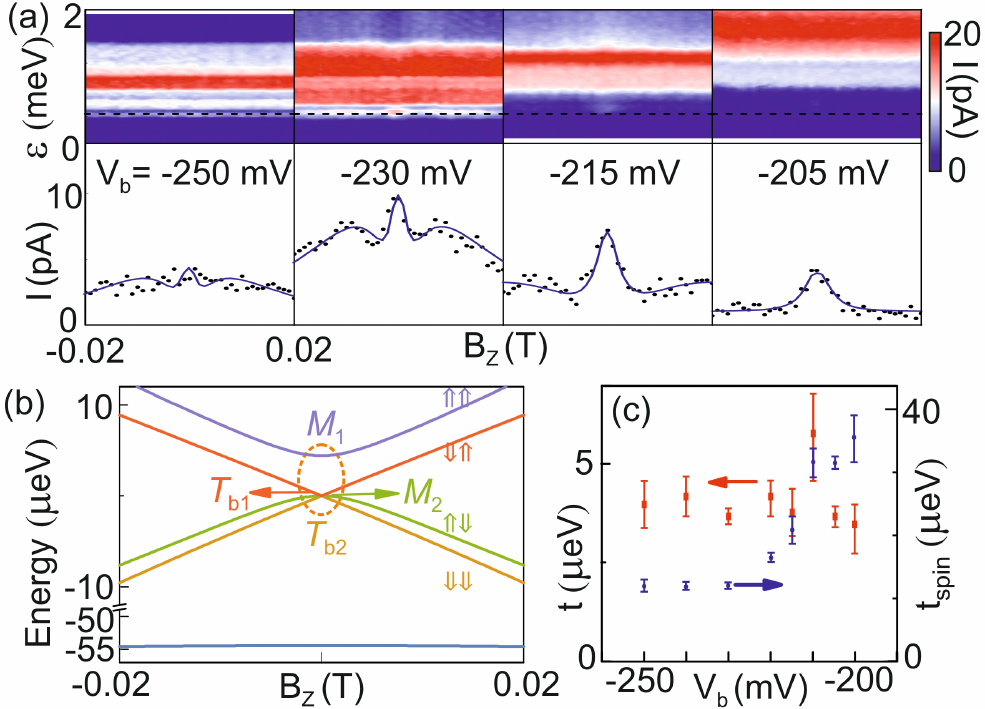}
	\caption{(a) Field-dependent transition measured with different tunnel barrier gate settings. Upper plots: current as a function of $\varepsilon$  and $B_Z$ in Pauli blockade. Lower plots: Cuts along dashed lines of constant detuning. Data (dots) are fitted by a model (curves) described in the text. (b) Schematic energy levels in the absence of hyperfine interaction. The zero-field current peak in (a) is associated with the level degeneracy at $B_Z=0$ (highlighted by ellipse), where hyperfine interaction mixes blocked and unblocked states. Side peaks are associated with spin-dependent tunneling. (c)~Tunnel coupling $t$ and $t_\text{spin}$ extracted from fits as in (a). Error bars represent 95~\% confidence intervals.}
	\label{Fig3}
\end{figure}
We now include hyperfine interaction, which acts on both spin and valley degrees of freedom~\cite{Palyi2009,Csiszar2014}.
At $\Bvec=0$,
the three energetically aligned states $\ket{T_{b1}}$, $\ket{T_{b2}}$, and
$\ket{M_2}$ mix to form new eigenstates
$\ket{M'_2}$, $\ket{M'_3}$, $\ket{M'_4}$,
each overlapping with $\ket{S}$ and therefore contributing to the current via spin-independent inelastic interdot tunnelling. In this picture, the triple-peak structure is explained as follows. Each current peak indicates a field strength where the $(n_h,1_e)$ eigenstates are singlet-triplet mixtures, allowing relaxation to $((n+1)_h,2_e)$.  The central peak arises from hyperfine mixing of three degenerate states [highlighted in Fig. 3(b)]. Side peaks are induced by the interplay of the Zeeman effect and effective spin-dependent interdot tunnelling.
At large detuning ($\varepsilon \gg t,\tspin $),
the energy scale characterizing spin mixing within the $(n_h,1_e)$
configuration is
$t_\text{spin} t / \varepsilon$ \cite{Supp}.
In general the preferred axis for spin-dependent tunneling aligns neither with the nanotube nor with $\Bvec$. In the field range where
$ t_\text{spin} t/\varepsilon \sim \mu_\text{B} B_Z$, the $(n_h,1_e)$ eigenstates are therefore singlet-triplet mixtures,
which results in side peaks in $\ISD(B_Z)$.
As $B_Z$ is further increased (such that $\mu_\text{B} B_Z \gg t_\text{spin} t /\varepsilon$), Zeeman energy dominates spin-orbit-induced mixing, so that the eigenstates are
$\ket{\Uparrow\Downarrow}$,
$\ket{\Downarrow\Uparrow}$,
$\ket{T_+}$, and
$\ket{T_-}$, where the latter two reestablish Pauli blockade.

We validate this picture quantitatively by fitting measured current [cuts in Fig. 3(a)] using a model of charge relaxation among the five spin-orbit and hyperfine mixed spin-valley states. Inelastic charge relaxation with rate $\Gammain$ causes
$(n_h,1_e)$ states to decay to $((n+1)_h, 2_e)$ based on their overlap with with $\ket{S}$. Nuclear-spin fluctuations are incorporated by averaging $I$ over an ensemble of hyperfine configurations~\cite{Supp} with root-mean-square coupling strength $\EN$.
We first fit the second panel usin fit parameters $\Gammain$, $t$, $t_\text{spin}$, and $E_\text{N}$. Holding the fitted value $E_\text{N}=0.16 \pm 0.03~\mu$eV, we then fit across the range of $\Vb$ settings. Fitted values of $t$ and $t_\text{spin}$ are shown in Fig.~3(c). Extracted $t$ is fairly constant over the range, whereas $t_\text{spin}$ increases with $\Vb$. This presumably reflects that whereas the interdot barrier of an n-p double dot is set by the slope of the potential and not strongly affected by $\Vb$, the Rashba spin-orbit coupling is set by the perpendicular electric field. Unexpectedly, we find $\tspin>t$.

\begin{figure}[t]
	\centering
	\includegraphics[width=1\columnwidth]{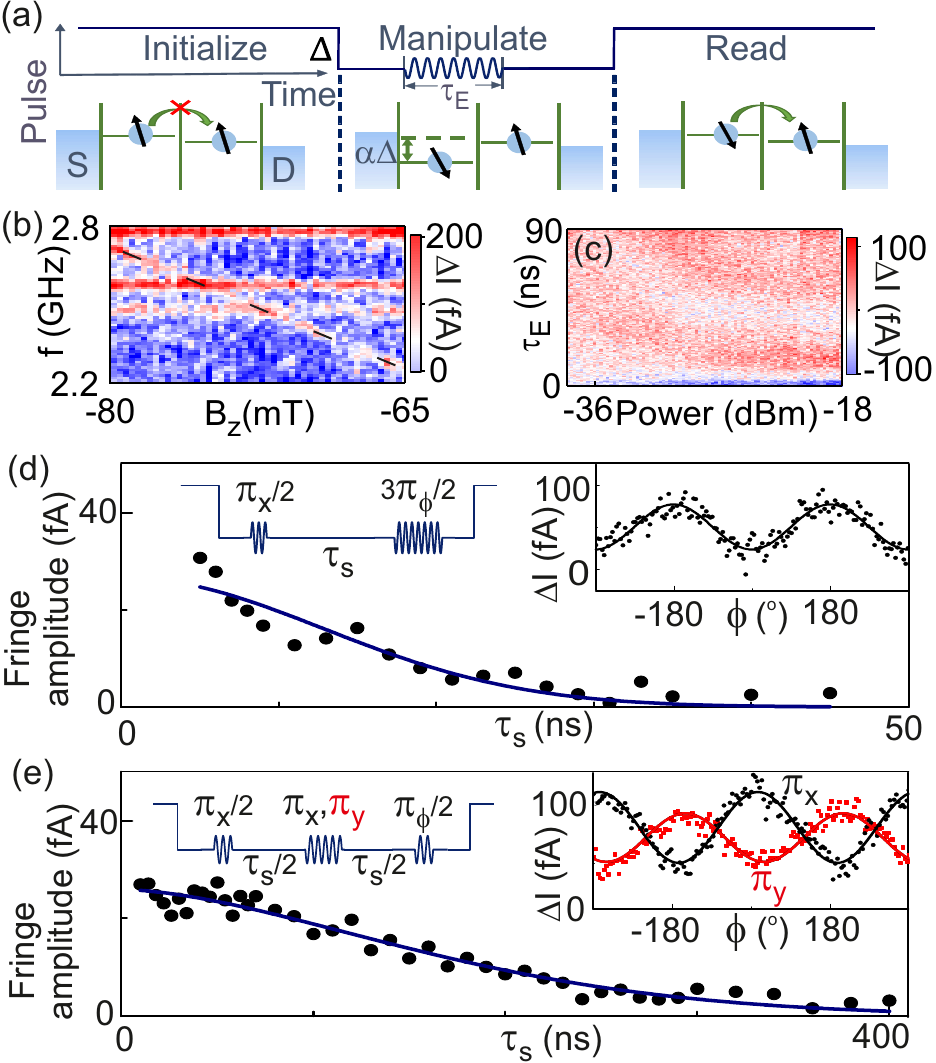}
	\caption{(a) Pulse scheme for qubit manipulation. Upper panel: gate voltage cycle applied to G1. Lower panel: double dot energy levels during each step. 
		(b) Resonance signal (marked by dashed line) as a function of $f$ and $B_Z$ with $\theta=0$, $\tauE=40$~ns, and microwave level $\sim -20$~dBm at the device. (c) Resonant signal as a function of $\tauE$ and of power at the device (d) Ramsey dephasing measurement. Points: Fringe amplitude as a function of $\tau_\text{S}$. Curve: Gaussian fit. Left inset: Ramsey pulse cycle. Right inset: signal as a function of phase difference with $\tau_\text{S} = 5$~ns. (e) Hahn echo measurement. Points: Fringe amplitude. Line: Fit (see text). Right inset: fringes for orthogonal phases of the echo pulse (along $x$ and $y$ axes in the qubit's rotating frame), with $\tauS = 10$~ns. In panels (c)-(e), $\theta = 15^{\circ}$, $B=83$~mT, and $f=2.82$~GHz}
	\label{Fig4}
\end{figure}

To further explore hyperfine interaction, we characterize a spin-valley qubit at this transition~\cite{Laird2013}. The qubit is controlled using electrically driven spin resonance (EDSR) with a cycle of gate voltage pulses applied to G1 and G5 [Fig.~4(a)] \cite{Koppens2006,Laird2007,Nowack2007}. The cycle first initializes an effective triplet state by configuring the double dot in Pauli blockade. The detuning is then pulsed to configure the device in Coulomb blockade, and a microwave burst at frequency $f$ is applied to G1 to manipulate the spin-valley state. Finally the device is returned to Pauli blockade; if an effective spin flip has occurred Pauli blockade is temporarily lifted, allowing the result of the manipulation to be read out via the current. Repeating the cycle with period $\sim 800$~ns, the resulting current change $\Delta \ISD$ is detected by chopping the microwaves at 117~Hz and locking in to the chopper signal~\cite{Laird2007}. The EDSR spectrum [Fig.~4(b)] shows a diagonal line of increased $\Delta \ISD$, indicating resonance when $f$ matches the qubit frequency $\fR$. The slope gives an effective parallel $g$-factor $g= 2.22 \pm 0.02$, which is nearly eight times smaller than the right-dot $g_\text{orb}$ extracted above from transport measurements but consistent with transport spectroscopy of the left dot~\cite{Supp}.

Qubit dephasing is measured using pulsed spectroscopy \cite{Koppens2006,Laird2013}. We operate at $|\mathbf{B}| = 83$~mT, $\theta = 15^\circ$, and $\fR=2.82$~GHz, which gives good contrast of the pulsed $\Delta\ISD$. (Previous experiments~\cite{Laird2013} found dephasing independent of $|\mathbf{B}|$ and $\theta$.) Applying a single microwave burst of duration $\tauE$ per pulse cycle drives coherent Rabi oscillations between qubit states [Fig.~4(c)]. As expected, Rabi frequency increases with microwave power, but saturates at the highest power suggesting a contribution of short-range disorder to the EDSR mechanism~\cite{Szechenyi2014}. With coherent manipulation established, we measure dephasing using a Ramsey sequence of two bursts per pulse cycle separated by time $\tauS$ [Fig.~4(d)]. As a function of phase difference $\phi$ between bursts, $\Delta \ISD$ shows fringes whose amplitude decays as $e^{-(\tau_S/\Tstar)^2}$, where $\Tstar$ is the dephasing time. A fit to the data gives $\Tstar= 13 \pm 1$~ns. Using a Hahn echo sequence [Fig.~4(e)] to cancel out slowly varying noise, the amplitude decays more slowly and is phenomenologically fit by $e^{(-\tau_\text{S}/T_\text{echo})^\gamma}$, with fitted coherence time $T_\text{echo} = 198 \pm 7$~ns and $\gamma =1.7 \pm 0.2$.

These values are similar to previous measurements on a spin-valley qubit~\cite{Laird2013}. That experiment did not allow  conclusive determination of dephasing or decoherence mechanisms, with charge noise~\cite{Mavalankar2016} and hyperfine coupling~\cite{Churchill2009Nat,Churchill2009PRL} being leading candidates. In our device, we now show that charge noise does not limit $\Tstar$. By changing the pulse voltage $\Delta$ for the manipulation step [Fig.~4(a)], we measure the dependence of $\fR$ on gate voltage [Fig.~5(a)]. A linear fit gives $d\fR/d\Delta = -0.4 \pm 0.3$~MHz/mV. (The orthogonal axis in gate space showed a similarly weak dependence~\cite{Supp}.) Thus to explain the measured $\Tstar$ by noise on the detuning axis would require root-mean-square voltage noise $\Deltarms \geq 27$~mV \cite{Taylor2007}. Since this is broader than the narrowest transport features this mechanism can be ruled out~\cite{Supp}. By a similar argument, the noise level to account for the measured $\Techo$ would be $\Deltarms \gtrsim 2$~mV~\cite{Taylor2007}. This is consistent with the data, although greater than the estimated instrument noise, implying an origin in the device itself. It is also approximately consistent (roughly six times larger) with an independent measurement of charge noise in a similar device~\cite{Mavalankar2016}. Temperature dependence of $\Tstar$ and $\Techo$ is shown Fig. 5(b).

\begin{figure}[t]
\centering
\includegraphics[width=1\columnwidth]{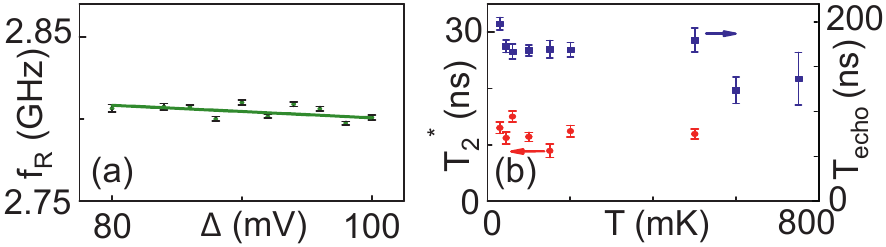}
\caption{(a) Qubit frequency as a function of pulse amplitude (points) with linear fit. (b) Measured $T_2^{\ast}$ ($\bullet$) and $T_\text{echo}$ (\protect\rule{1ex}{1ex}) as a function of temperature. Above 750~mK, incoherent current leakage prevents EDSR measurement.} 
\label{Fig5}
\end{figure}

In conclusion, both leakage current and qubit dephasing imply hyperfine coupling to a randomly fluctuating spin bath of $^{13}$C nuclei in each quantum dot, with effective coupling strength $\EN \sim 0.16~\mu$eV. Considering the estimated $6\times 10^4$ nuclei in each dot and 1.1~\% $^{13}$C abundance, this implies hyperfine constant $\mathcal{A} \sim 4 \times 10^{-4}$~eV. This is consistent with other measurements on isotopically purified~\cite{Churchill2009PRL} and natural~\cite{Laird2013} nanotube devices, but continues a long-standing discrepancy with numerical simulations~\cite{Yazyev2008,Fischer2009,Csiszar2014}, and bulk spectroscopy of fullerenes~\cite{Pennington1996nuclear} and nanotubes~\cite{Ihara2010,kiss2011enhanced}. Hyperfine interaction may also limit $\Techo$, but since the measured value implies unexpectedly rapid nuclear spin diffusion~\cite{Supp}, we suspect that charge noise is more significant. This would indicate the spin-valley qubit is sensitive to electric fields, for example because of interdot exchange~\cite{Laird2013,Li2014}.

We acknowledge Templeton World Charity Foundation, EPSRC (EP/J015067/1), Marie Curie CIG and IEF fellowships, Stiftung der Deutschen Wirtschaft, and the Royal Academy of Engineering.

\newpage

\clearpage
\setcounter{equation}{0}
\setcounter{figure}{0}
\setcounter{table}{0}
\makeatletter

\onecolumngrid
\parbox[c]{\textwidth}{\protect \centering \Large \MakeUppercase{Supplementary Material}}
\rule{\textwidth}{1pt}
\vspace{1cm}
\twocolumngrid

\section{Nanotube synthesis}

Synthesis of the nanotube begins by dissolving FeCl$_3$ in a solution of PMMA in anisol. The solution is spun onto a quartz chip in which pillars 4.5~$\mu$m tall have been etched. The chip is heated in a chemical vapor deposition furnace to $900^{\circ}$C and exposed to a 20:80 H$_2$:Ar atmosphere to reduce the FeCl$_3$ to Fe catalyst. Nanotubes are then grown at 950$^{\circ}$C from a 20:80 CH$_4$:Ar atmosphere.

\section{Single-dot Coulomb spectroscopy}

\begin{figure}[t]
	\centering
	\includegraphics[width=0.8 \columnwidth]{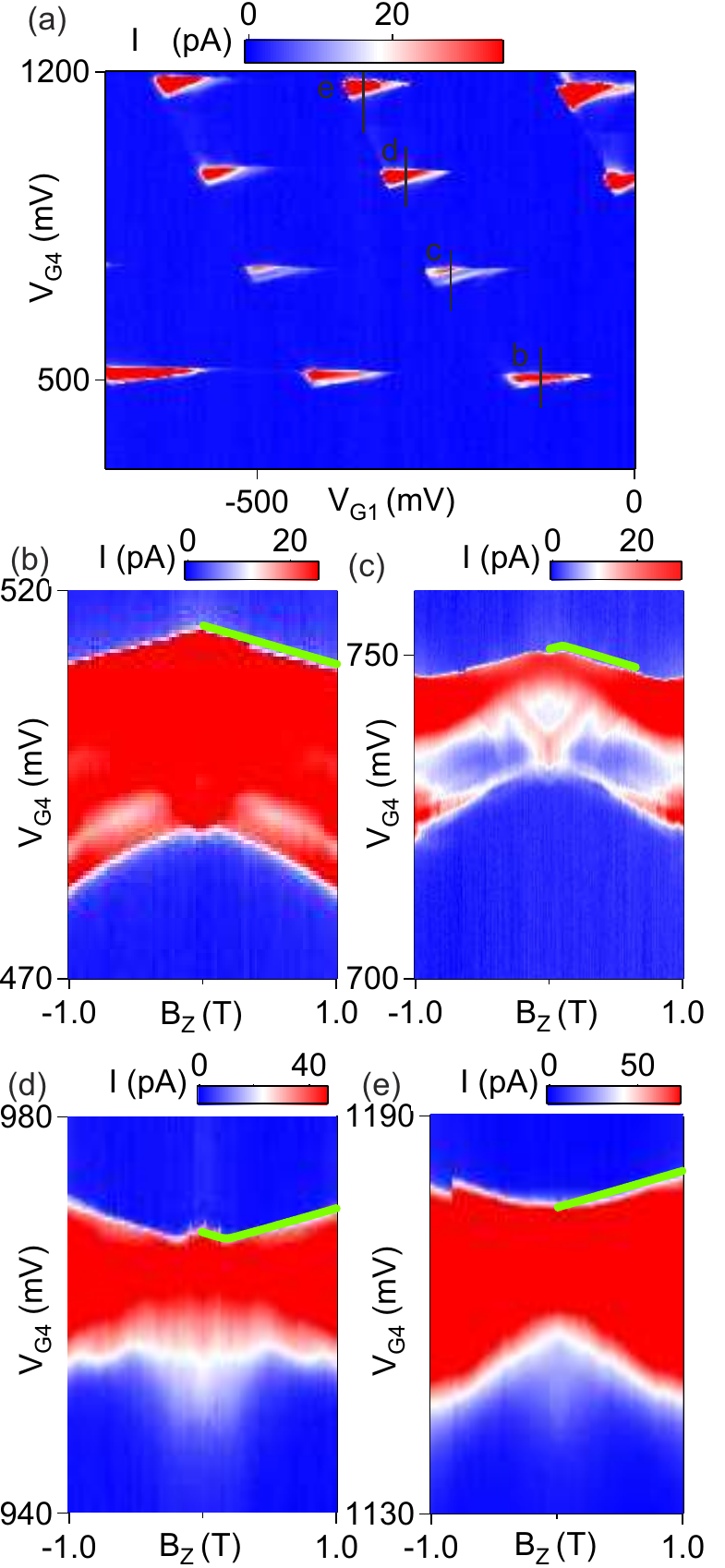}
	\caption{(a) Current as a function of $V_\text{G1}$ and $V_\text{G4}$ with $\VSD=10$~mV, for a slightly different gate setting to Fig.~1. The four marked bias triangles correspond to the first four electron transitions of the right quantum dot in the p-n regime. (b)-(e) Current as a function of $V_\text{G4}$ and $B_Z$, measured along the black lines marked in panel (a). Green lines run parallel to each upper edge, which maps out the ground-state energy of the right dot.}
	\label{FigS1}
\end{figure}

\begin{figure}[t]
	\centering
	\includegraphics[width=0.8\columnwidth]{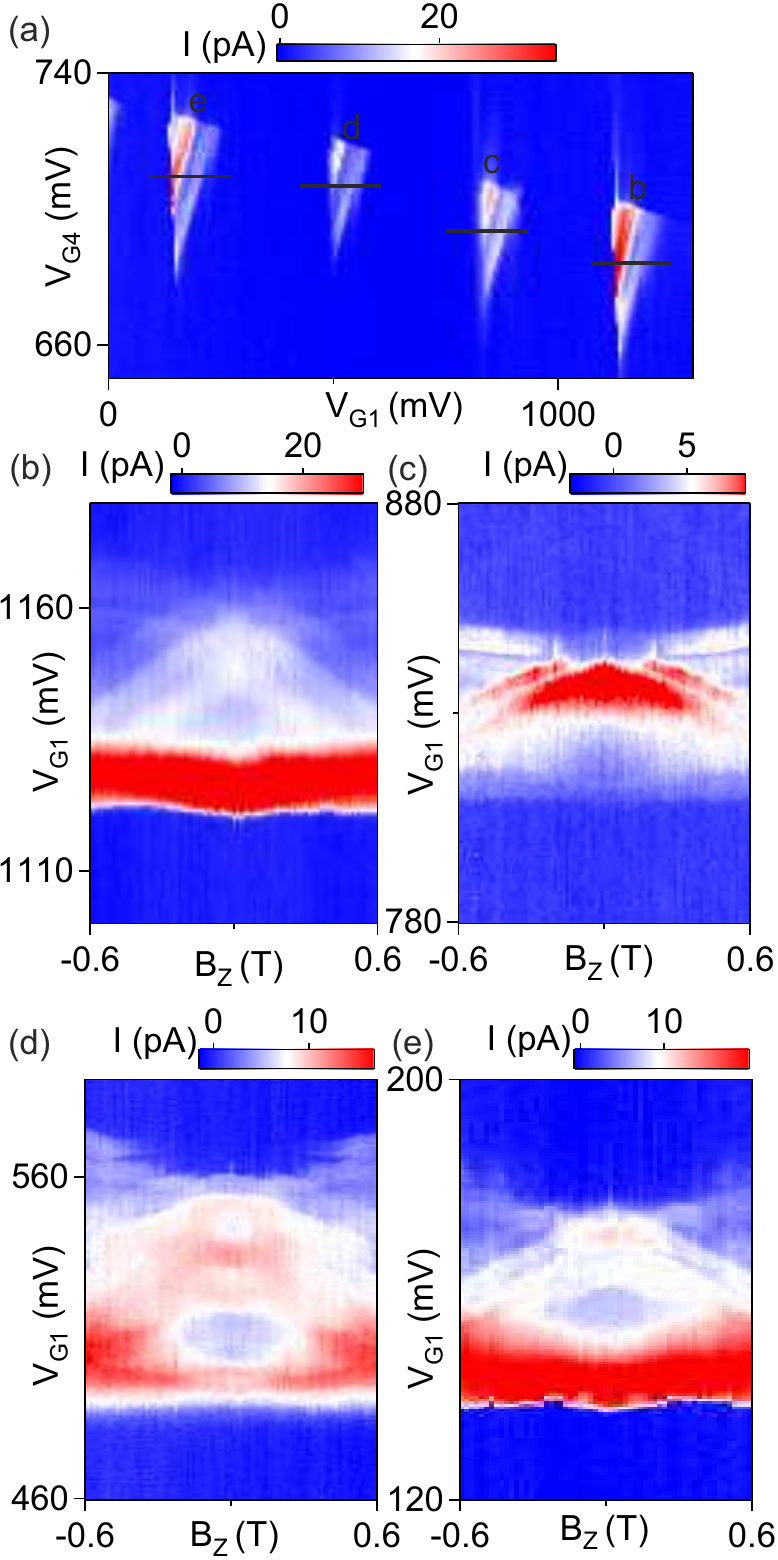}
	\caption{(a) Current as a function of $V_\text{G1}$ and $V_\text{G4}$ with $\VSD=10$~mV, for a different gate setting from that of Fig.~1. The four marked bias triangles correspond to four successive hole transitions of the left quantum dot in the p-n regime. (b)-(e) Current as a function of $V_\text{G4}$ and $B_Z$, measured along the black lines marked in panel (a). The lower edges track the ground-state energy of the left dot.}
	\label{FigS2}
\end{figure}

In this section we show transport spectroscopy of the single-quantum dot energy levels on left and right. Figure~S1(a) shows the double-dot stability diagram in a region of gate space corresponding to a p-n configuration, with gate settings similar to Fig.~1 of the main text. The first four bias triangles in one column, indicating transitions $(n_h,m_e)\rightarrow((n+1)_h,(m+1)_e)$ for $m=0,1,2,3$, are selected to study the energy spectrum of the right dot. In Fig.~S1(b-e), current in these triangles is shown as function of $B_Z$ and $V_\text{G4}$ along the lines marked in Fig.~S1(a). The upper edge of each triangle marks the degeneracy between the ground-state chemical potential of the right dot and the right lead, and therefore measuring this edge location as a function of $B_Z$ reveals the energy spectrum of the right dot~\cite{Pei2012}. From the measured spectra for the first four holes, we extract for the nanotube spin-orbit coupling $\Delta_\text{SO}=450$ $\mu$eV, for the valley mixing parameter $\DeltaKK\le 80$ $\mu$eV and for the orbital $g$-factor $g_\text{orb}=19$ \cite{Kuemmeth2008,Steele2013,Laird2015}. These values are similar to the measurements from Fig.~2 of the main text.

Figure S2 shows similar measurements of the left quantum dot for a nearby set of transitions, showing a sequence of bias triangles corresponding to transitions $(n_h,1_e)\rightarrow((n+1)_h,2_e)$ for four successive values of $n_h$. As in Fig.~S1, transitions marked by black lines in are measured as a function of $B_Z$ and $V_\text{G1}$, allowing the ground-state energy of the left dot to be extracted from the lower edge of each bias triangle. Because the left dot could not be depleted by accessible gate voltages, the absolute number of holes is not known. In contrast to the right dot, the left dot does not show an enhanced parallel $g$-factor, or any clear sign of spin-orbit coupling. We ascribe this to suppression of the orbital $g$-factor by the large number of holes in the left dot, or by intervalley scattering \cite{Jespersen2011}.

\section{Leakage current in a transverse
magnetic field}

In the main text, Fig.~2(c) shows the leakage current
$I(B_X)$ through the device in a transverse magnetic field $B_X$.
The current resonance at zero detuning, that is, the uppermost white
horizontal stripe at $V_\text{G1} \approx 940\, \text{mV}$ in
Fig.~2(c), shows no appreciable dependence
on $B_X$ in a wide magnetic-field window
$B_X \in [-0.8,0.8] \, \text{T}$.
This is in stark contrast to the zero-field peaks observed in
a longitudinal field $B_Z$ (Fig.~2(b) and 3(a)). In this section we provide a qualitative explanation, referencing quantitative theory~\cite{Szechenyi2013}.

Reference~\onlinecite{Szechenyi2013} shows how
such a sustained leakage current in a transverse field
can arise if the complex-valued
 valley-mixing matrix elements in the two  dots,
$\Delta_{KK'}^{(L)} = |\Delta_{KK'}^{(L)}| e^{i\varphi_L}$
and
$\Delta_{KK'}^{(R)} = |\Delta_{KK'}^{(R)}| e^{i\varphi_R}$,
are different.
See, for example, Fig.~5(b) and (e) in Ref.~\onlinecite{Szechenyi2013}, which
show the field dependence of the leakage current
when the current flow is dominated by
inelastic $(1,1) \to (0,2)$ tunneling processes,
and the difference of the valley-mixing phases is $\varphi_L - \varphi_R = \pi/2$.

The reason for the sustained
leakage current is as follows.
The homogeneous external magnetic field
$(B_X,0,0)$
induces effective magnetic (Zeeman) fields
$(\mathcal B_1, \mathcal B_2,0)$,
acting on each low-energy
effective spin-$1/2$ Kramers doublet
[see Eq.~(5) of Ref.~\onlinecite{Szechenyi2013}].
This effective Zeeman field
has two important
properties [see Eq.~(6a) and (6b) of Ref.~\onlinecite{Szechenyi2013}]:
(i) it is rotated around the third axis by the valley-mixing angle $\varphi$;
(ii) its magnitude is scaled by the absolute value
$|\Delta_{KK'}|$ of the local valley-mixing matrix element.
The properties (i) and (ii) imply that the effective Zeeman fields in the
two dots are different in magnitude and direction, leading to singlet-triplet mixing and a finite leakage current~\cite{Jouravlev2006}. By contrast, a longitudinal magnetic field induces parallel effective fields in the two dots, which does not by itself lead to singlet-triplet mixing.

Over a wide range in $B_X$, this leakage current is governed by the relative directions (not by the magnitudes) of the effective Zeeman fields [see Eq.~(11) of  Ref. \onlinecite{Jouravlev2006} and Eq.~(24) of Ref.~\cite{Szechenyi2013}]. The current is therefore independent of $B_X$. This holds so long as the effective Zeeman splitting dominates the exchange energy but is less than the detuning:
\begin{equation}
\label{eq:fieldrange}
t^2/\varepsilon \lesssim \mu_{B} |\mathcal{B}| \lesssim \varepsilon.
\end{equation}
For small $B_X$, such that the first inequality in Eq.~(\ref{eq:fieldrange}) is violated, this leakage mechanism is ineffective. . However, hyperfine interaction induces a significant leakage current for small $B_X$. Therefore we conclude that the $B_X$ dependence of the leakage current is characterized by a zero-field value $I_\mathrm{hf}$ set by hyperfine interaction, and a $B_X$-independent value in a wide range of $B_X$, $I_\mathrm{vm}$, set by valley mixing.

\begin{figure}[t]
\centering
\includegraphics[width=1\columnwidth]{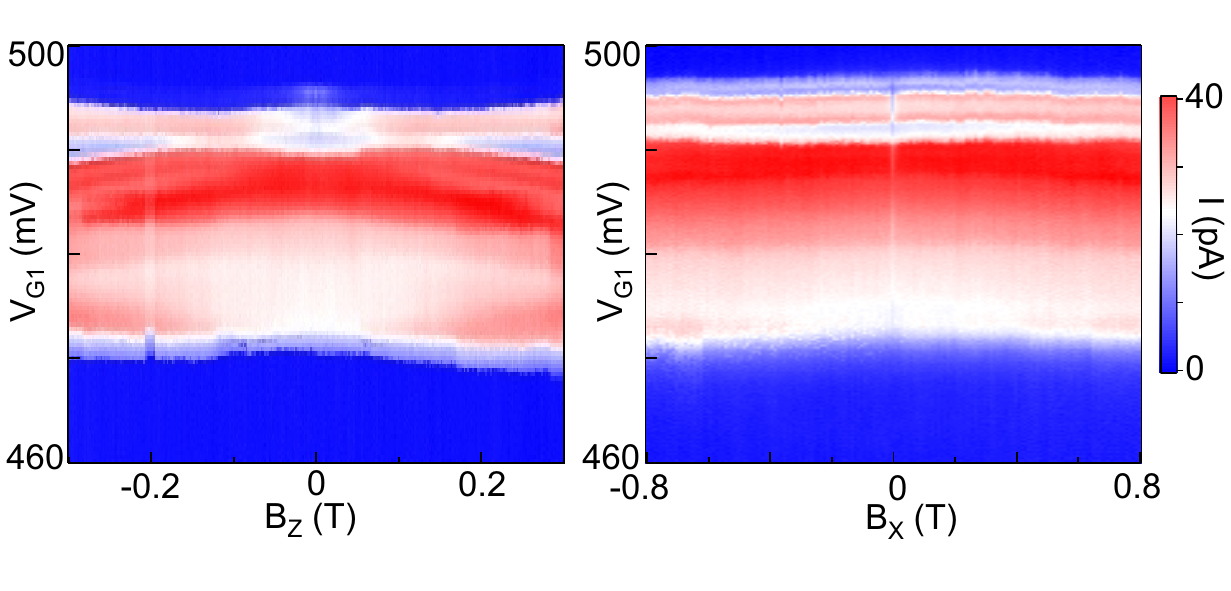}
\caption{Current at a blocked transition as a function of $V_\text{G1}$ along the detuning axis and of magnetic field. In longitudinal magnetic field (a), the current shows a zero-field peak due to hyperfine mixing; in transverse field (b) there is a zero-field dip and a sustained $B_X$-independent current for finite $B_X$, consistent with the presence of different valley-mixing phases in the two dots.}
\label{FigS3}
\end{figure}

In general, $I_\mathrm{hf}$ and $I_\mathrm{vm}$ are different, and their relation depends on the valley mixing matrix elements, which in turn depend on the electronic orbitals participating in the transport process. Therefore, the relation of $I_\mathrm{hf}$ and $I_\mathrm{vm}$ can be different at different charge transitions in a given device. Fig.~2(c) in the main text shows a case when the zero-field $I_\mathrm{hf}$ and finite-field $I_\mathrm{vm}$ current values are experimentally indistinguishable. Fig.~\ref{FigS3}, however, shows data corresponding to a different charge transition in the same device, where the $B_X$ dependence of the leakage current shows a zero-field dip (Fig.~\ref{FigS3}(b)), i.e., the zero-field current $I_\mathrm{hf}$ is significantly lower than the finite-field current $I_\mathrm{vm}$.

Related anisotropic Pauli blockade was previously seen in nanotubes in Ref.~\cite{Pei2012}, in nanowires consisting of InAs \cite{schroer2011}, InSb \cite{Nadj-Perge2012}, and SiGe \cite{Brauns2016}, and in planar silicon devices \cite{li2015pauli}. This work is the first to measure and explain the distinctive triple peak in leakage current (Fig. 3(a)) that arises from the interplay of hyperfine and spin-orbit interaqction.

\section{Leakage current in a longitudinal magnetic field}

\subsection{The Hamiltonian}
This section gives details of the model used to generate the fits in Fig.~3.
The model is based on Refs. \onlinecite{Jouravlev2006} and~\onlinecite{Danon2009}.
It is sufficient to focus  on the five-dimensional two-electron Hilbert space
representing the (1,1) and (0,2) charge configurations.
Here the four basis states of the (1,1) charge configuration,
$\ket{S}$,
$\ket{T_-}$,
$\ket{T_0}$, and
$\ket{T_+}$,
are defined in the usual way from the low-energy
Kramers-pair
single-electron states $\{\ket{\Uparrow}, \ket{\Downarrow}\}$ in left and right
dots, and the fifth basis state $\ket{S_g}$ corresponds
to the (0,2) charge configuration.
The Hamiltonian is:
\bean
H = H_\text{d} + H_\text{tun}+H_{B} + H_\text{hf},
\eean
where the terms on the right- hand side represent the (1,1)-(0,2) energy detuning,
spin-dependent interdot tunneling, coupling to the external magnetic field,
and hyperfine interaction, respectively.
These terms read
\bean
H_\text{d} &=& -\varepsilon \ket{S_g}\bra{S_g},\\
\label{eq:interdot}
H_\text{tun} &=&
t \ket{S}\bra{S_g}  +
i t_\text{spin} \vec{n} \cdot \ket{\vec T} \bra{S_g}
+ \text{h.c.}, \\
H_B &=& \frac 1 2 \mu_B B_Z \left(
	g_L  \sigma_{Lz} + g_R  \sigma_{Rz}
\right),
\\
H_\text{hf} &=&
\vec B_{NL} \cdot \vec \sigma_L +
\vec B_{NR} \cdot \vec \sigma_R.
\eean
As in the main text, $\varepsilon$ is the (1,1)-(0,2) energy detuning, and
$t$ and $t_\text{spin}$ are respectively the spin-independent and spin-dependent tunnel amplitudes.
In Eq.~(\ref{eq:interdot}), we have defined a vector of triplet states $\ket{\vec T} = (\ket{T_x},\ket{T_y},\ket{T_z})$,
with
$\ket{T_{x}} = (\ket{T_-} - \ket{T_+})/\sqrt{2}$,
$\ket{T_{y}} = i (\ket{T_-} + T_+)/\sqrt{2}$,
and
$\ket{T_z} = \ket{T_0}$,
in order to characterize the spatial direction of the spin-dependent tunneling term~\cite{Danon2009} with a unit vector $\vec n$.
This interdot tunneling Hamiltonian $H_\text{tun}$ is the same as introduced
in the main text, with the identification $\vec n \cdot \ket{\vec T}
= \ket{T_u}$.
Effective Zeeman coupling is parameterized by effective $g$-factors $g_L$, $g_R$ in left and right dots, where $\vec \sigma_{Lz}$ and $\vec \sigma_{Rz}$ are the Pauli effective spin operators and $\mu_B$ is the Bohr magneton. In contrast to Ref.~\onlinecite{Nadj-Perge2010}, we do not assume equal $g$-factors in the two dots. Hyperfine interaction is parameterized by effective nuclear magnetic fields $\vec B_{NL}$, $\vec B_{NR}$ (with dimensions of energy) in the two dots.

\subsection{Spectrum and tunnel rates}
For simplicity, we consider the special case
that the spatial direction $\vec n$ characterizing the
spin-dependent part of the interdot tunneling is perpendicular to $Z$.
This is the case, e.g. if the spin-dependent tunneling is
induced by Rashba spin-orbit interaction~\cite{Klinovaja2011} arising from
an electric field perpendicular to $Z$.
We choose $\vec n = (0,1,0)$.

We consider large detuning,
$\varepsilon \gg t, t_\text{spin}, g_{L,R}\, \mu_B |B_Z|, E_\text{N}$,
where $E_\text{N}$ is the characteristic energy scale
of the hyperfine interaction.
We numerically diagonalize $H$ to obtain the energy eigenstates.
In this large-detuning condition,
four of the energy eigenstates have most of their
weight in the (1,1) charge configuration;
we denote these as $\ket{j}$ with $j = 1,2,3,4$.
The fifth eigenstate has most of its weight in the (0,2) charge
configuration; we denote this as $\ket{\tilde S_g}$.

\begin{figure*}
	\centering
	\includegraphics[width=170 mm]{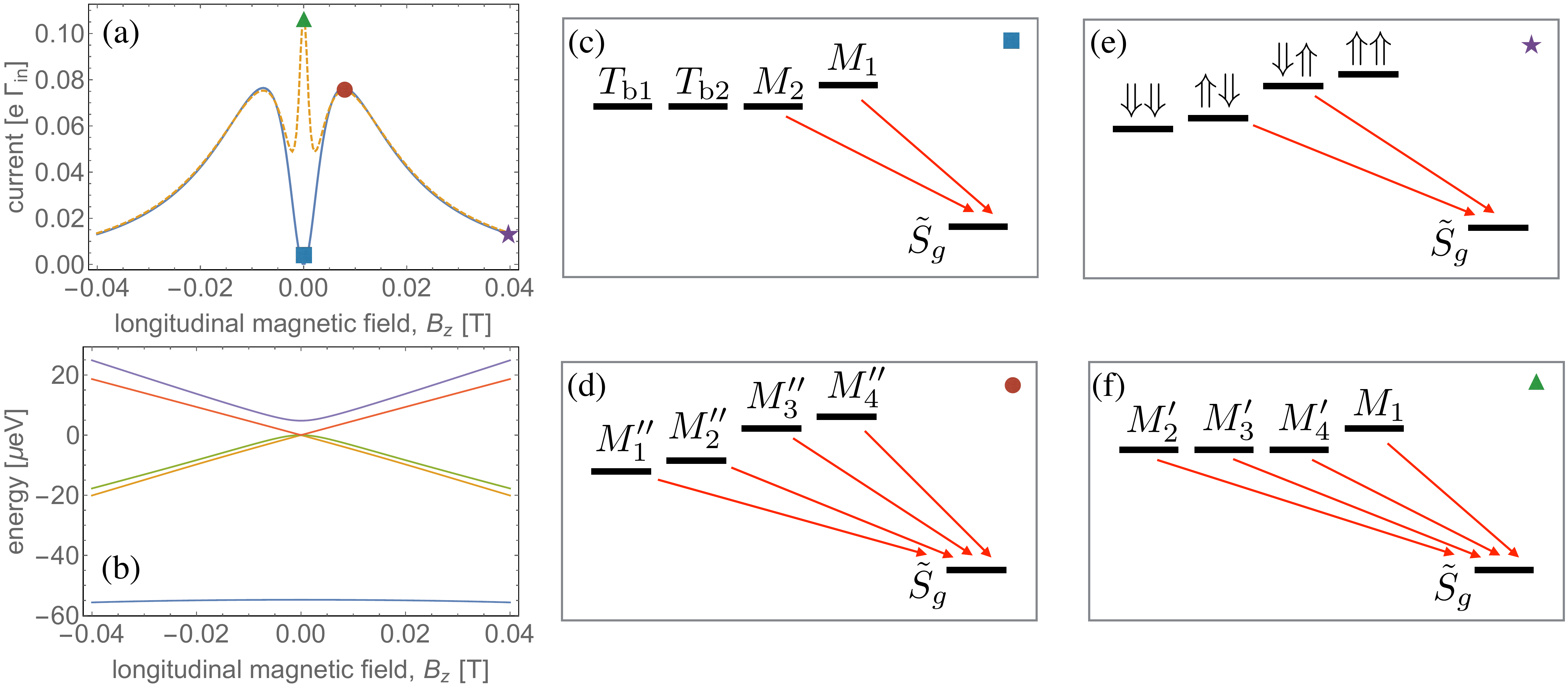}
	\vspace{-0.3cm}
	\caption{
		Leakage current, energy spectrum, and inelastic downhill
		interdot tunneling processes in a longitudinal magnetic field.
		(a) Simulated leakage current as a function of magnetic field without (solid line, $\EN=0$) and with (dashed line, $\EN=0.16 \, \mu$eV) hyperfine interaction.
		Other parameters:
		$t=3 \, \mu$eV,
		$t_\text{spin}=11\, \mu$eV,
		$\varepsilon = 50 \, \mu$eV,
		$g_L = 2$, $g_R = 17$;
		number of nuclear-field realizations: 5000.
		(b) Energy eigenvalues of the tunnel-coupled (1,1)-(0,2) states
		as functions of the longitudinal magnetic field $B_Z$. Parameters as in (a) wth $\EN=0$.
		(c,d,e,f) Energy levels (black horizontal lines) and inelastic,
		energetically downhill interdot tunneling processes (red arrows)
		for the parameters marked in (a).
	}
	\label{fig:withouthyperfine}
\end{figure*}

Adopting the physical picture leading to Eq.~(10) of Ref.~\onlinecite{Jouravlev2006},
from now we assume that current is carried by
spin-independent inelastic relaxation
from the (1,1) configuration to the (0,2) configuration, for example by phonon-assisted tunneling.
We denote the rate characterizing such a
transition from $\ket{S}$ to $\ket{S_g}$ by $\Gamma_\text{in}$.
Then the transition rate from the energy eigenstate $\ket{j}$ to $\ket{\tilde S_g}$
is
\bean
\Gamma_{\tilde{S}_g \leftarrow j} &=&
\Gamma_\text{in} \left|
	\braket{S| j}
\right|^2.
\eean
The current is calculated from the harmonic mean of transition rates~\cite{Jouravlev2006},
\bean
\ISD = 4 e \left(
	\sum_{j=1}^4 \Gamma_{\tilde{S}_g \leftarrow j}^{-1}
\right)^{-1}.
\label{eq:Itotal}
\eean
Following Ref.~\onlinecite{Jouravlev2006}, the non-static nature of the
nuclear spins is taken into account by averaging
the computed current over a large number of random nuclear-field
realizations.
For a given realization, each Cartesian component of
$\vec B_{NL}$ and $\vec B_{NR}$ is
drawn from a Gaussian ensemble with zero mean and
standard deviation $E_\text{N}$. The fits in Fig.~3 are derived from numerical simulations of Eq.~(\ref{eq:Itotal}) over 5000 nuclear field realizations, with $\Gammain$, $t$, $\tspin$, and $\EN$ as input parameters.

\subsection{Magnetic field dependence}

In the context of this model, we now provide an interpretation of the triple-peak pattern
in the longitudinal-field magnetocurrent $\ISD(B_Z)$,
observed for some settings of the barrier gate voltage $V_\text{b}$
(e.g. second column, bottom row of Fig.~3(a) in the main text).
Within this interpretation, the central peak around $B_Z = 0$
is induced by the interplay of hyperfine interaction and
spin-dependent interdot tunneling,
whereas the side peaks are caused by the latter effect alone.

First, we consider the situation with no hyperfine interaction ($E_\text{N} = 0$). The simulated current in this case is shown by a solid line in Fig.~\ref{fig:withouthyperfine}(a).
Further parameter values are given in the caption.
The current is zero for $B_Z = 0$,
shows two side peaks around
$B_Z \approx \pm 5 $~mT,
and decays as $B_Z$ is further increased.

At $B_Z = 0$, the current is zero for the following reason.
The Hamiltonian in this case consists
of two terms only, detuning and tunneling,
$H = H_\text{d} + H_\text{tun}$.
Since we took $\vec n = (0,1,0)$, the Hamiltonian
$H$ leaves the two triplet states $\ket{T_x}$ and $\ket{T_z}$
uncoupled from the singlets (Fig.~3(c)).
Therefore these two states,
denoted $\ket{T_\text{b1}}$ and
$\ket{T_\text{b2}}$ in the main text and in
Fig.~\ref{fig:withouthyperfine}(c),
block the current.
The other three basis states, $\ket{S}$, $\ket{S_g}$ and $\ket{T_y}$,
are mixed by the coherent spin-independent and spin-dependent
interdot tunneling process described by $H_\text{tun}$.
At large detuning, one of the energy eigenstates,
$\ket{\tilde{S}_g}$, stays predominantly in the (0,2) charge configuration,
whereas the remaining two energy eigenstates $\ket{M_1}$ and
$\ket{M_2}$ have predominantly (1,1) character. The states
$\ket{M_1}$ and $\ket{M_2}$ are mixtures of $\ket{S}$ and $\ket{T_y}$, hence both
states can relax to $\ket{\tilde{S}_g}$.
This is indicated by the red arrows in Fig.~\ref{fig:withouthyperfine}(b).
The fact that the leakage current is zero for $B_Z=0$ does not rely on the specific choice of $\vec n$:
the form of $H_\text{tun}$ guarantees that for any
$\vec n$, there is only one state in the triplet
subspace (namely, $\vec n \cdot \ket{\vec T}$)
which is mixed with the singlets.

For intermediate $B_Z$, when the Zeeman splittings
are comparable to the energy scale of the spin mixing
caused by spin-dependent interdot tunneling,
the interplay of $H_\text{tun}$ and $H_B$
can result in an efficient mixing between $\ket{S}$ and all three triplet states (Fig.~\ref{fig:withouthyperfine}(d)).
The four resulting energy eigenstates $\ket{M''_1}$, $\ket{M''_2}$, $\ket{M''_3}$
$\ket{M''_4}$ all have finite $\ket{S}$ component, and therefore
can relax to $\ket{\tilde{S}_g}$.
This leads to the broad side peaks seen in Fig.~\ref{fig:withouthyperfine}(a).
An important condition for efficient singlet-triplet mixing and therefore
non-zero leakage current is that  the spin-orbit direction $\vec n$
is not aligned with $\vec B$.

An estimate for the position of the side
peak can be obtained using quasidegenerate
perturbation theory \cite{Winkler2003},
relying on the large-detuning condition.
In the basis $\{S_g$,
$\Downarrow \Downarrow$,
$\Downarrow \Uparrow$,
$\Uparrow \Downarrow$,
$\Uparrow \Uparrow\}$,
the Hamiltonian $H=H_\text{d}+H_\text{tun}+H_B$ reads
\begin{widetext}
\bean
H=\left(
\begin{array}{ccccc}
 -\varepsilon  & t_\text{spin} & -t & t & t_\text{spin} \\
 t_\text{spin} & -\frac{1}{2} \mu_B B_Z (g_L+g_R) & 0 & 0 & 0 \\
 -t & 0 & -\frac{1}{2} \mu_B B_Z (g_L-g_R) & 0 & 0 \\
 t & 0 & 0 & \frac{1}{2} \mu_B B_Z (g_L-g_R) & 0 \\
 t_\text{spin} & 0 & 0 & 0 & \frac{1}{2} \mu_B B_Z (g_L+g_R)
\end{array}
\right).
\eean
\end{widetext}
From this $H$, using second-order quasidegenerate perturbation theory in the small parameters $t^2/\varepsilon$, $t t_\text{spin}/\varepsilon$ and $t^2_\text{spin}/\varepsilon$, the following effective Hamiltonian is obtained for
the (1,1) subspace spanned by
$\{\Downarrow \Downarrow$,
$\Downarrow \Uparrow$,
$\Uparrow \Downarrow$,
$\Uparrow \Uparrow\}$:
\begin{widetext}
\bean\label{eq:hprime}
\tilde{H} =
\left(
\begin{array}{cccc}
 \frac{t_{\text{spin}}^2}{\varepsilon }-\frac{1}{2} \mu_B B_Z (g_L+g_R) & -\frac{t t_{\text{spin}}}{\varepsilon } & \frac{t t_{\text{spin}}}{\varepsilon } & \frac{t_{\text{spin}}^2}{\epsilon } \\[1.2ex]
 -\frac{t t_{\text{spin}}}{\varepsilon } & \frac{t^2}{\varepsilon }-\frac{1}{2} \mu_B B_Z (g_L-g_R) & -\frac{t^2}{\varepsilon } & -\frac{t t_{\text{spin}}}{\varepsilon } \\[1.2ex]
 \frac{t t_{\text{spin}}}{\varepsilon } & -\frac{t^2}{\varepsilon } & \frac{t^2}{\varepsilon }+\frac{1}{2} \mu_B B_Z (g_L-g_R) & \frac{t t_{\text{spin}}}{\varepsilon } \\[1.2ex]
 \frac{t_{\text{spin}}^2}{\varepsilon } & -\frac{t t_{\text{spin}}}{\varepsilon } & \frac{t t_{\text{spin}}}{\varepsilon } & \frac{t_{\text{spin}}^2}{\varepsilon }+\frac{1}{2} \mu_B B_Z (g_L+g_R)
\end{array}
\right).
\eean
\end{widetext}
Consider now the $2\times 2$ subblock of $\tilde{H}$ corresponding to
$\{\Downarrow \Uparrow, \Uparrow \Uparrow\}$.
At sufficiently large positive $B_Z$,
when $g_R \mu_B B_Z \gg t_\text{spin}^2/\varepsilon,\,
t t_\text{spin}/\varepsilon, \, t^2/\varepsilon$,
these two states are energetically well separated
from the other two states by the Zeeman splitting
because $g_L \ll g_R$.
With the parameters of Fig.~\ref{fig:withouthyperfine}, this condition is
$B_Z \gg 5.2$ mT, cf. the  purple and red lines in
Fig.~\ref{fig:withouthyperfine}(b).
The energy splitting between
$\ket{\Downarrow \Uparrow}$ and $\ket{\Uparrow \Uparrow}$
is $\frac{t^2_\text{spin}-t^2 }{\varepsilon} + g_L \mu_B B_Z$,
which evaluates to $\sim 2.8 \, \mu$eV at
$B_Z = 5.2$ mT.
As seen from $\tilde{H}$ in Eq.~\eqref{eq:hprime},
the same two states are mixed by the term
$\frac{t t_\text{spin}}{\varepsilon} \approx 0.66 \, \mu$eV.
Thus the degree of mixing between
the triplet state $\ket{\Uparrow \Uparrow}$
and the other basis states becomes progressively weaker
as the magnetic field is increased above 5.2 mT,
implying that the leakage current also becomes
more and more suppressed (Fig.~\ref{fig:withouthyperfine}(e)).
This prediction based on quasidegenerate perturbation theory compares well
with the trend in the
numerical simulation (Fig.~\ref{fig:withouthyperfine}(a)).

Now considering the situation with hyperfine interaction, the leakage current
develops an additional zero-field peak,
as shown by the dashed orange
line in Fig.~\ref{fig:withouthyperfine}(a).
As discussed in the main text, the reason is
as follows.
Hyperfine interaction in carbon nanotubes acts on both the spin and
valley degrees of freedom \cite{Palyi2009,Csiszar2014}.
Therefore it mixes
the three energetically aligned states $\ket{T_{b1}}$, $\ket{T_{b2}}$ and
$\ket{M_2}$, resulting in new eigenstates
$\ket{M'_2}$, $\ket{M'_3}$, $\ket{M'_4}$ (Fig.~\ref{fig:withouthyperfine}(f)),
all of which have non-zero overlap with $\ket{S}$ and therefore
contribute to the current.

\subsection{Details of fit procedure}
Fits in Fig.~3(a) are run using Matlab's \texttt{lsqcurvefit} routine, which performs nonlinear least-squares fitting. The 95\% confidence intervals in Fig.~3(c) are extracted from these fits using the \texttt{nlparci} routine.

\section{Quantitative discussion of dephasing and decoherence mechanisms}
In this section, we numerically estimate, based on our experimental results, the strength of effects that limit $\Tstar$ (dephasing) and $\Techo$ (decoherence).

\subsection{Detuning noise}
First, we calculate the possible contribution of detuning noise. Our model is that the qubit frequency $f$ during the manipulation pulse depends on the detuning $\Delta$, and that $\Delta$ is subject to noise with a one-sided power spectral density $S_{\Delta\Delta}(F)$, where $F$ is frequency parameter.

\subsubsection{Contribution to dephasing}
\label{sec:dephasing}
With root-mean-square detuning jitter $\Delta_\mathrm{rms}$, the distribution of detuning values over successive repetitions of the EDSR burst cycle is
\begin{equation}
P(\Delta) = \frac{1}{\sqrt{2\pi}\Deltarms}e^{-\Delta^2/2\Deltarms^2},
\end{equation}
where $\Deltarms$ includes fluctuations up to a frequency $F\sim 1/\Tstar$.
The signal measured in Fig.~4, which is the average over many repetitions, is therefore proportional to the correlator
\begin{align}
C(\tauS) 	&= \int_{-\infty}^\infty P(\Delta) \cos \left(2\pi\Delta \frac{df}{d\Delta} \tauS\right) d\Delta\\
		&= e^{-\Deltarms^2\tauS^2/2|d\Delta/df|},
\end{align}
giving $\Tstar=1/(\sqrt{2}\pi\Deltarms|df/d\Delta|)$. From the measured $|df/d\Delta|\geq0.7$~MHz/mV, we therefore conclude that the amount of detuning noise needed to explain the measured $\Tstar \geq 12$~ns would be
\begin{equation}
\Deltarms\geq 27~\mathrm{mV}.
\end{equation}
Since the narrowest features measured in DC transport are $\sim 3$ mV wide, we therefore exclude charge noise as the origin of the short $\Tstar$.

\subsubsection{Contribution to decoherence}
 The effect on qubit coherence was calculated in Ref.~\onlinecite{Taylor2007}; the qubit state decays as:
\bean
C(\tauS) = \exp\left(-\frac{\langle \Xi^2(t)\rangle}{2}\right),
\eean
where $C(\tauS)$ is the qubit correlator and
\begin{equation}
\langle \Xi^2(\tauS)\rangle\equiv2\pi^2~\mathrm{Re}\int_0^\infty dF~S_{ff}(F) \frac{\sin^2 \pi F\tauS/2}{(\pi F/2)^2} [1-e^{i\pi F\tauS}].
\label{eq:xi}
\end{equation}
Here
\bean
S_{ff}(F)=\left(\frac{df}{d\Delta}\right)^2 S_{\Delta\Delta}(F),
\eean
is the power spectral density of the jitter of the qubit frequency.
Since low-frequency components of $S_{\Delta\Delta}(F)$ are weakly weighted in the integral Eq.~(\ref{eq:xi}), reflecting the fact that they are cancelled in the echo sequence, the contribution of low-frequency noise to limiting $T_\text{echo}$  is weak. (These low-frequency components do of course limit $\Tstar$.) We therefore consider as a model that limits $T_\text{echo}$  efficiently for a given total noise power that $S_{\Delta\Delta}(F)$ is independent of $F$ (i.e. white) over a frequency range extending from zero to a few times  $1/T_\text{echo}$. Equation~(\ref{eq:xi}) then gives
\begin{align}
\langle \Xi^2(\tauS)\rangle	&= 2\pi^2  S_{ff}~\mathrm{Re}\int_0^\infty dF \frac{\sin^2 (\pi F\tauS/2)}{(\pi F/2)^2} [1-e^{i\pi F\tauS t}] \\
					&= 8\pi S_{ff}\tauS ~\mathrm{Re}\int_0^\infty dx \frac{\sin^2 x}{x^2} [1-e^{2ix}] \\
					&= 4 \pi^2 S_{ff} \tauS.
\end{align}

In other words, white noise gives exponential decay $C(\tauS)=\exp (-\tauS/T_\text{echo})$ with \\
\begin{align}
T_\text{echo} 	&= \frac{1}{4\pi^2 S_{ff}}\\
			&= \frac{1}{4\pi^2 (df/d\Delta)^2 S_{\Delta\Delta}}.
\end{align}
From our measured $T_\text{echo} \approx 200$ ns, we thus predict that the detuning voltage noise level needed to give the observed decoherence is
\bean
\sqrt{S_{\Delta\Delta}} &\geq& \frac{1}{2\pi |df/d\Delta|\sqrt{T_\text{echo}}} \\
&\approx& 0.5~\mu\mathrm{V}/\sqrt{\mathrm{Hz}}.
\eean
Integrated over a frequency range up to $F_\text{cutoff}  \sim 2/T_\text{echo}$, this requires a total root-mean-square detuning jitter
\bean
\Delta_\text{rms} \geq \sqrt{S_{\Delta\Delta}F_\text{cutoff}} \sim 2~ \text{mV}
\end{eqnarray}
to explain the measured $\Techo$.

\subsection{Hyperfine coupling}
Here we estimate a hyperfine coupling from the Pauli blockade leakage current, as performed previously in an enriched $^{13}$C device~\cite{Churchill2009NAT} and in a natural-abundance device~\cite{Pei2012}. With hyperfine coupling $\mathcal{A}$, the hyperfine energy is $E_\text{N} \sim \mathcal{A} \sqrt{f_{13}/N}$, where  $f_{13}=1.1 \%$ is the isotopic fraction of $^{13}$C and $N$ is the number of atoms in each quantum dot. Assuming a single-walled nanotube of diameter $D\sim4$ nm and dot length $\sim 70$ nm, we have $N \sim 6 \times 10^4$ carbon atoms. The fitted $\EN =  0.16$ therefore implies $\mathcal{A} \sim 4 \times 10^{-4}$ eV, consistent with previous estimates for quantum dots \cite{Churchill2009NAT, Pei2012}. As previously pointed out, these estimates of $\mathcal A$ are apparently inconsistent with simulations \cite{Yazyev2008} and EPR measurements \cite{Pennington1996nuclear}.

This hyperfine coupling is expected to limit $\Tstar$ to $\sim\hbar/\EN = 4.1$~ns, of the same order of the measured value, suggesting that hyperfine interaction in the main limit on the measured value of $\Tstar$. The coherence time $T_\text{echo}$ can in principle also be limited by hyperfine interaction because nuclear spin diffusion gives rise to evolution of the effective hyperfine field between the two halves of the echo sequence. Assuming that nuclear spin diffusion gives rise to fluctuations with correlation time $T_\text{corr}$, the resulting coherence time\cite{Taylor2007} is $T_\text{echo} \sim 8^{1/4}\sqrt{T_2^{\ast}T_\text{corr}}$. To explain our measurements, this time would have to be $T_\text{corr} \sim 1~\mu$s. This is surprisingly fast compared with measurements in e.g. GaAs quantum dots.

In conclusion, our results imply $\Tstar$ is limited by hyperfine interaction, but the limit for $\Techo$ cannot be securely attributed either to hyperfine interaction or to charge noise.

\section{Dephasing due to voltage noise on other gates}

\begin{figure}
\centering
\includegraphics[width=1\columnwidth]{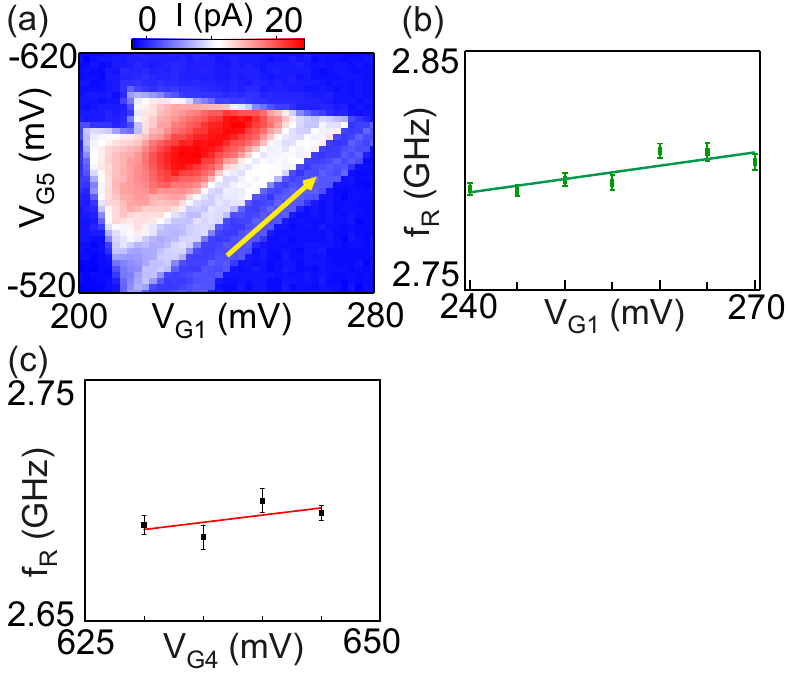}
\caption{(a) Current of the blockade transition as a function of $V_\text{G1}$ and $V_\text{G5}$ with $\VSD=8$~mV and $\Vb=-210$~mV.(d) The dependence of $\fR$ on gate voltage along the yellow arrow. (c) Resonance frequency measured with different voltage applied to $V_\text{G4}$, when $B_{Z}$ fixed at 0.079 T.  }
\label{FigS5}
\end{figure}

Section~\ref{sec:dephasing} showed that voltage noise along the detuning axis does not cause the observed qubit dephasing. Here we present analogous measurements for other axes in gate voltage space. Figure~\ref{FigS5}(b) shows the dependence of qubit frequency on gate voltage along an orthogonal axis in $\{V_\text{G1},V_\text{G5}\}$ space [Fig.~\ref{FigS5}(a)] . A linear fit gives slope $d\fR/dV_\text{G1} = -0.6 \pm 0.2$~MHz/mV. According to Eq. (S13), to explain $\Tstar \geq 12$~ns, we need $\Deltarms\geq 27$~mV. However, even if we attribute the thermal broadening of the Coulomb peak measured as a function of $V_\text{G4}$ all to charge noise, it only gives us $\Deltarms\leq 2$~mV.

We also consider noise coupling to other gates G2-G4. Unlike G1 and G5, these gates are not coupled to high-frequency lines and are much better filtered at low temperature~(with two-pole 100~kHz $RC$ filters~\cite{Mavalankar2016}). However, these gates affect inter-dot tunneling more than G1 and G5.
Figure~\ref{FigS5}(c) shows the dependence of $f_R$ on G4, from which we extract $d\fR/dV_\text{G4} = 0.6 \pm 0.5$~MHz/mV. Thus to explain the measured $\Tstar$ would require charge noise on G4 of $\Delta_\mathrm{rms}\geq 17~\mathrm{mV}$ which is again larger than the narrowest measured transition. We did not measure $\fR$ as a function of G2 and G3; however, since the gate pitch is almost the same with suspended nanotube height, we do not expect drastically different capacitive coupling from their neighboring gates.


\vfill

%


\end{document}